# Evaluating Generative AI Tools for Personalized Offline Recommendations: A Comparative Study

Rafael Salinas
Computer Science Department
University of Cuenca
Cuenca, Ecuador
rafael.salinasb@ucuenca.edu.ec

Otto Parra
Computer Science Department
University of Cuenca
Cuenca, Ecuador
otto.parra@ucuenca.edu.ec

Nelly Condori-Fernández
Computer Science Department
Universidad Santiago de Compostela
Santiago de Compostela, Spain
n.condori.fernandez@usc.es

Maria Fernanda Granda
Computer Science Department
University of Cuenca
Cuenca, Ecuador
fernanda.granda@ucuenca.edu.ec

*Abstract*—[Background] Generative AI tools have become increasingly relevant in supporting personalized recommendations across various domains. However, their effectiveness in health-related behavioral interventions, especially those aiming to reduce the use of technology, remains underexplored. [Aims] This study evaluates the performance and user satisfaction of the five most widely used generative AI tools when recommending non-digital activities tailored to individuals at risk of repetitive strain injury. [Method] Following the Goal/Question/Metric (GQM) paradigm, this proposed experiment involves generative AI tools that suggest offline activities based on predefined user profiles and intervention scenarios. The evaluation is focused on quantitative performance (precision, recall, F1-score and MCC-score) and qualitative aspects (user satisfaction and perceived recommendation relevance). Two research questions were defined: RQ1 assessed which tool delivers the most accurate recommendations, and RQ2 evaluated how tool choice influences user satisfaction.

## I. INTRODUCTION

Generative Artificial Intelligence (GenAI[1]) tools present a promising opportunity to deliver personalized, context-aware offline activity recommendations [1]. However, little is known about how different GenAI systems perform in this context, particularly regarding their suggestions' relevance and user satisfaction.

Recommender systems are essential tools on digital platforms, helping users find relevant products, services, or content based on their preferences. While they improve user engagement and satisfaction, such as in e-commerce and streaming services, their growing complexity often leaves users unaware of how personalized suggestions are generated [2].

Repetitive Strain Injury (RSI) describes a set of disorders resulting from repeated motions, excessive use, or extended periods of activity that put stress on muscles, tendons, and nerves, especially in areas like the hands, wrists, arms, shoulders, and neck. It typically impacts people who engage in repetitive actions, such as keyboard typing, mouse use, or performing routine tasks on production lines [3].

Software developers often work long hours in front of screens, significantly increasing their risk of developing RSI [4], [5], [6]. Traditional preventive strategies, such as static ergonomic checklists [7], automated break reminders [8], or wellness programs, are typically generic, reactive rather than proactive, and frequently disregarded by users due to lack of personalization and engagement [9], [10].

Furthermore, recent studies in the field of software engineering highlight the rising importance of developer well-being. For instance, the AI can help reduce developer stress by automating repetitive and mentally demanding tasks, like code review, debugging, and task prioritization, permitting the developers to focus on creative problem-solving[11]. Similarly, one of the latest systematic reviews on software developers' well-being reveals that it can be understood as a complex, multi-dimensional construct. Based on personal traits and workplace conditions, well-being can be predicted. Enhancing well-being leads to improved performance and healthier work environments [12].

While recent research has increasingly focused on the application of GenAI tools to enhance software development productivity, there is a notable gap in understanding their potential to support developers' physical and emotional well-being [13]. Building upon previous work on persuasive and context-aware recommendations using a predefined message catalogue [14], this study takes advantage of the approach by leveraging large language models (LLMs) to automatically generate tailored activity suggestions, enabling broader personalization and scalability. To address this gap, the use of LLMs as digital interventions is explored to: (i) Recommend personalized, health-preserving activities tailored to developers at risk of RSI. (ii) Evaluate usability and emotional engagement through a combination of self-reported feedback and affective computing techniques, such as emotion recognition.

This research addresses the main question: "*How do different generative AI tools compare recommendation performance and user satisfaction when providing personalized offline activity suggestions to mitigate repetitive strain injury (RSI) in subjects playing the role of software*

---

[1] In this work the terms "GenAI", "LLMs" and "generative AI tools" are considered similar.

*developers?*". To the best of current knowledge, this represents the first comparative evaluation of LLMs applied to offline, health-related activity recommendations for RSI prevention.

This research contributes to the growing field of human-centered software engineering by emphasizing the importance of supporting the physical and emotional well-being of software developers through intelligent, adaptive interventions. The contribution of this paper is twofold: (i) a comparative analysis of generative AI tools for producing non-digital activity recommendations that support developer well-being, and (ii) the identification of measurable differences in the effectiveness and user perception of these tools, highlighting key trade-offs between system performance and user satisfaction.

This paper is structured as follows: In section II, the theoretical background is included to give readers a better comprehension. Section III, Related Work, includes relevant topics identified and analyzed based on a literature review. Section IV contains the Methodology used in this research work. Section V describes the results obtained in the evaluation process and their data analysis. Section VI discusses threats of validity. Section VII discusses the ethical issues involved in the research.

## II. THEORETICAL BACKGROUND

This section provides foundational background information crucial for understanding the context and methodology of this study.

GenAI refers to recent advances in Artificial Intelligence, particularly in LLMs. Generative models have been proven valuable in different fields, including software engineering, by enabling innovative ways to perform various tasks, such as auto-completing code [15].

A brief description of each GenAI tool used in this study can be seen briefly in TABLE I. Following each description:

Gemini: Gemini is Google's family of multimodal AI models for advanced reasoning, image and text processing, and code generation. It integrates capabilities from DeepMind's AlphaCode and is optimized for web-scale applications and integration with Google tools [16].

Phi-4: It is part of Microsoft's line of lightweight, high-performance language models aimed at efficiency and low resource usage. Despite its small size, it is optimized for educational, and reasoning tasks and performs surprisingly well on benchmark tests [17].

Mistral: It is an open-weight language model designed for versatility and speed. It uses a dense transformer architecture and is known for its strong performance in multilingual and code-related tasks, with models like Mistral 7B and Mixtral gaining popularity in the open-source community [18].

Qwen 2.5: it is the latest version of the Qwen models from Alibaba, known for their strong multilingual support, high performance in coding and instruction following, and integration into various enterprise applications across Asia [19].

LLaMA 3.2: LLaMA (Large Language Model Meta AI) 3.2 refers to a possible future or variant of the LLaMA 3 series. LLaMA 3 models are designed for high-quality language generation and are trained with open-weight principles, aiming to compete with the best proprietary models in performance and transparency [20].

TABLE I. SUMMARIZING MODELS COMPARED

| Model | Developer | Open Source | Optimized For | Notable Feature |
|---|---|---|---|---|
| Gemini | Google DeepMind | No | Multimodal + Code | Deep integration with Google |
| Phi-4 | Microsoft | Yes | Low-resource, reasoning | Small footprint |
| Mistral | Mistral AI | Yes | Speed, multilingual | Dense transformer |
| Qwen 2.5 | Alibaba DAMO | Yes | Instruction following | Multilingual focus |
| LLaMA 3.2 | Meta | Yes | General-purpose language | Open-weight, privacy focus |

Repetitive Strain Injury (RSI), on the other hand, is a set of musculoskeletal disorders resulting from repeated motions or prolonged use of computers and mobile devices, especially among younger individuals with lifelong exposure to technology [3], [5], [6]. Based on these preliminary studies, software developers are at high risk due to long hours of keyboard and mouse use. Traditional preventive strategies, such as static ergonomic checklists [7], automated break reminders [8], or wellness programs, are typically generic, reactive rather than proactive, and frequently ignored by users due to a lack of personalization and engagement [9], [10]. Some technological solutions have been developed to address this issue. Tools like WorkRave, XWrits, Stretch Break, and RSIGuard remind users to take short breaks and stretch at regular intervals, but are often ignored because they disrupt workflow or lack relevance to the user's context [21].

From a theoretical rationale point of view aligned with personalized offline recommendations, recent research in ergonomics and occupational health has shown that short, frequent breaks not only help prevent injury but can also improve productivity, emphasizing the need for more attractive and proactive ergonomic interventions [22], [23]. However, the effectiveness of such interventions is closely linked to their personalization and contextual relevance. According to recent systematic reviews, well-being among software developers is a complex, multi-dimensional construct influenced by both personal traits and workplace conditions [12]. Therefore, personalized interventions that account for individual preferences, schedules, and contexts are more likely to be adopted and effective.

Since recommender systems are essential tools in digital platforms, helping users find relevant products, services, or content based on their preferences, in the health domain, recommender systems have been used to promote healthy behaviors. However, most existing solutions are generic and lack the ability to adapt to individual needs in real time [24]. GenAI tools, particularly LLMs, offer a new opportunity to generate highly personalized, context-aware offline activity recommendations. By leveraging detailed user profiles—including work schedules, preferred activity types, and contextual factors such as time of day and weather—LLMs can generate recommendations that are not only tailored to the user's needs and interests but also dynamically adapted to their real-world context. This approach is theoretically expected to enhance user engagement and adherence, ultimately reducing RSI risk and supporting software developer well-being.

## III. RELATED WORK

Recently, there has been a notable increase in the utilization of Generative AI techniques in recommender systems. Two works that summarize what is found in the literature are described below.

Ayemowa et al. [24] describe in a systematic literature review generative AI models, particularly Generative Adversarial Networks (GANs) and Variational Autoencoders (VAEs), that are gaining attention in healthcare recommender systems. These models offer improved personalization, handling of sparse data, and the ability to simulate user behavior, key features for applications like treatment suggestions, health risk predictions, and therapy personalization. The study calls for future research to explore deeper integration of generative AI in healthcare applications, emphasizing the need for ethical frameworks, data protection, and interdisciplinary collaboration between AI developers and medical professionals.

Said [2] presents a systematic literature review on using Large Language Models (LLMs), such as LLaMA and ChatGPT, to enhance the explainability of recommender systems. The study analyses current approaches, outlines key challenges, and proposes future research directions. It highlights the promise of LLMs in creating more transparent and user-friendly recommendation explanations.

Several studies addressing RSI and Musculoskeletal Disorders (MSDs) have highlighted the critical role of workplace ergonomics in mitigating the risks associated with these conditions. The study presented in [25] proposed a machine learning framework that generates a detailed prompt based on posture sequence predictions and associated uncertainty estimates. This prompt is subsequently processed through an API by a LLM, such as GPT-4 or LLaMA-2, to produce an interpretable occupational health risk assessment and tailored user recommendations. Nevertheless, the scenario described in that study does not adequately reflect the specific context and needs of the software developer population.

The SuperBreak software aims to increase break compliance by offering interactive activities during breaks that make them less intrusive and more productive. However, its personalization remains limited. It primarily considers basic preference-based settings, the user's preferred interaction style (active vs. passive), a single self-reported form regarding ergonomic concerns and work habits (without tracking user history), and some sensitivity to the office environment. A more comprehensive approach to user profiling would likely enhance the effectiveness of such interventions. Moreover, this continues to rely on technology to perform RSI prevention activities [21].

A complementary approach was proposed in [14], where a persuasive and context-aware recommendation framework to promote RSI prevention through offline physical activities was developed. The system used a fixed catalogue of manually designed suggestions and considered contextual cues such as user interruptibility and persuasive message framing. Although the approach enabled more tailored and timely interventions compared to fixed reminder systems, it still relied on manually authored content. In contrast, the present study explores whether LLMs can generate activity suggestions that are both relevant and emotionally engaging for users and compares the outputs of multiple LLMs in terms of quality and user satisfaction.

To the present authors' knowledge, no previous reports have existed about using recommender systems for personalized offline recommendations to mitigate RSI. This research will evaluate five GenAI tools to apply in this area. In addition, unlike related work, which typically relies on static preference settings or limited user inputs, the approach of this study leverages a significantly more detailed user profile to generate personalized offline activity suggestions. Specifically, class and work schedules are incorporated, users' preferred types of non-technological activities (e.g., physical, artistic, social, cultural, relaxation, creative, or educational), and contextual factors such as day of the week, time of day, and weather conditions. This richer input allows for the generation of recommendations that are not only tailored to the user's needs and interests, but also dynamically adapted to their real-world context, enhancing both relevance and user engagement.

## IV. METHODOLOGY

### A. Evaluation design

This study compares five widely used GenAI models to analyze their ability to generate relevant, personalized offline activity recommendations based on predefined user profiles and scenarios.

In accordance with the Goal/Question/Metric (GQM) Paradigm [26], [27], the goal of this empirical study follows the schema in TABLE II.

TABLE II. GQM SCHEMA

| Goal | Question | Metrics |
|---|---|---|
| Analyze the performance of LLMs in generating relevant offline activity recommendations for users at risk of RSI. | Which generative AI tool provides the most relevant recommendations for offline activities in RSI? | - Precision<br>- Recall<br>- F1-score<br>- MCC-score |
| Evaluate user satisfaction with the AI-generated offline activity recommendations. | How does the choice of a GenAI tool affect user satisfaction with the recommendations? | - User Satisfaction<br>- Emotional Response |

To analyze the recommendations generated by each GenAI model, the performance metrics will be computed regarding recommendation relevance, treated as a classification task. It is necessary to define a clear ground truth and a consistent evaluation logic. Since the goal is to assess generative AI tools for recommending non-digital activities tailored to software developers at risk of RSI, a user profile–driven scenario will be established.

Each recommendation will be evaluated based on the following criteria: (i) Relevance to the user's physical condition, as determined by a health professional, (ii) the Appropriateness to the user's demographics, habits, and preferences, assessing whether the recommendation aligns with time and space constraints, and (iii) Usefulness, as determined by user feedback.

In this context, a "Relevant Recommendation" is classified using: (i) Expert-labeled datasets: An Occupational Health Professional evaluates each recommendation according to the user profile. (ii) User feedback: Users assess whether the recommendations are applicable to their situation and how they feel about following the suggested instructions. Based on these inputs, recommendations will be quantified with a binary score: 1 ("relevant") or 0 ("irrelevant"). A final score is then calculated as the average of the previous two indicators (expert evaluation and user feedback). To ensure consistency in the binary classification of recommendation relevance, inter-rater agreement between expert evaluations will be measured using Cohen's Kappa coefficient [28]. Then, to address this goal, from the main research question and, based on the PICO framework[29], two research questions are defined, which are stated in TABLE III.

RQ1: **Which generative AI tool provides the most relevant recommendations for offline activities in RSI?** This RQ aims to determine which GenAI tool provides the most relevant recommendations to help software developers suffering from RSI.

RQ2: **How does the choice of a GenAI tool affect user satisfaction with the recommendations provided?** This RQ aims to define how user satisfaction is affected depending on the GenAI tool selected.

TABLE III. RESEARCH QUESTIONS PROPOSED

| Description | Research Question 1 | Research Question 2 |
|---|---|---|
| P(Population) | Software developers with risk or symptoms of RSI | Software developers with risk or symptoms of RSI |
| I(Intervention) | Use of a specific GenAI tool | Offline recommendations generated by one GenAI tool |
| C(Comparison) | Other GenAI tools used for the same task | Recommendations from other GenAI tools |
| O(Outcome) | Relevance of offline recommendations. | User satisfaction: SUS score, emotional state. |

To answer RQ1, four performance metrics commonly reported for evaluating the performance of LLMs are applied:

- **Precision (M1):** Proportion of relevant activities among all recommended ones [30].
- **Recall (M2):** Proportion of relevant activities correctly recommended [30].
- **F1-score (M3):** Harmonic mean of precision and recall [30].
- **MCC-Matthews Correlation Coefficient- (M4):** it considers true positives (TP), true negatives (TN), false positives (FP), and false negatives (FN), making it a more balanced and informative metric, especially in binary classification which aligns well with this study [31].

For RQ2, another metric is proposed:

- **User Satisfaction (M5):** To evaluate user satisfaction with each AI-generated recommendation, a set of emotional responses is identified to reflect the user's affective experience during interaction. Specifically, the following emotions are analyzed: happiness, anger, disgust, fear, sadness, and surprise[32]. Prior work has shown that both positive and negative aspects are central to evaluating satisfaction in human-product interactions [33]. Emotional cues will be extracted from participants' facial expressions recorded during interaction. Additionally, after reviewing all recommendations, the System Usability Scale (SUS) will be adapted to assess the perceived usefulness and clarity of the recommendations [34].

ISO 9241 defines User Satisfaction as freedom from discomfort and positive attitudes towards the generated recommendations [35], which in this study is operationalized through both emotional responses and usability ratings.

The main independent variable in this study is the Generative AI technology, operationalized through the specific LLM used to generate each recommendation: Gemini, Phi-4, Mistral, Qwen 2.5, and LLaMA 3.2. Other factors that might influence are:

- Potential bias in LLM-generated recommendations: To address this, prompt engineering was used to refine queries, and expert manual review ensured the recommendations were accurate, relevant, and aligned with professional judgment.

- Order of presenting recommendations: To minimize potential order effects, the sequence in which recommendations from each LLM are shown will be randomized.

- History of physical discomfort or RSI: A shortened version of the Nordic Musculoskeletal Questionnaire (NMQ), adapted to the Spanish language, will be incorporated into this form to generate a comprehensive user profile [36].

- Prior experience with AI: To rate participants' familiarity with AI tools, a single-item 5-point scale will be used: How would you rate your familiarity with AI-powered tools or virtual assistants?

The hypotheses proposed in this research work are:

$H_{11}$: At least one generative AI tool will perform better, as measured by precision, recall, F1-score, and MCC score

$H_{10}$: No generative AI tool performs better than the others; all tools have equal precision, recall, F1-score and MCC-score.

$H_{21}$: Users will be more satisfied with the recommendations generated by the model with the highest F1-score or MCC-score.

$H_{20}$: Users are not satisfied with the recommendations generated by the model with the highest F1-score or MCC-score; user satisfaction is equal across models regardless of these metrics.

The experiment first establishes a ground truth set of expert-defined activities and computes precision, recall, F1score and MCC-score for each LLM; a statistical comparison tests $H_{10}$ vs. $H_{11}$, rejecting $H_{10}$ if any model achieves a significantly higher value on either F1-score or MCC-score. It then collects participant feedback on the clarity, relevance, and usefulness of the recommendations, measures satisfaction via the SUS scale and facial emotion analysis, and tests $H_{20}$ vs. $H_{21}$ by examining whether the LLM with the highest, either F1score or MCC-score, also yields significantly higher user satisfaction scores.

The potential trade-offs between the selected performance metrics are acknowledged. For instance, one model may achieve higher precision, while another may excel in recall. To address this, the F1-score is defined as the primary comparative metric, as it provides a harmonic balance between precision and recall [37]. Additionally, some studies have raised concerns about the reliability of this metric in different scenarios [38]. Therefore, the metric M4, is proposed to support a more unbiased and exploratory analysis. A model is considered to perform significantly better if it achieves the highest F1-score or MCC-score among the tested models, supported by statistical analysis.

To carry out this process, first a normality test to each evaluation metric will be applied. If the data are normally distributed, an ANOVA test will be used to compare the performance (precision, recall, F1-score and MCC) and user satisfaction (SUS scores) across the five LLMs. If the data are not normally distributed, a Friedman test will be used. Significant results will be followed by post-hoc pairwise comparisons with appropriate correction for multiple testing. Effect sizes will be reported to quantify the magnitude of observed differences. This statistical approach will allow determining whether any LLM significantly outperforms others in terms of recommendation quality and user satisfaction.

### B. Participants and Procedures

The present study involved a convenience sample of 80 final-year Computer Science students from the University of Cuenca (Ecuador). All participants were at least 18 years old, owned a mobile phone, and volunteered to take part in the research. As a preliminary phase, a first round of interviews was conducted with these students, considering their active involvement as software developers in various academic and professional projects.

Participants with programming and coding backgrounds share core tasks, work environments, and ergonomic risks common to software developers in industry [39]. Therefore, insights gathered from this group can be reasonably generalized to the broader software engineering population. Their firsthand experience enables them to provide meaningful feedback on the relevance and usability of the recommendations, making this study a valuable starting point for more comprehensive research that spans diverse roles and levels within software engineering.

A demographic questionnaire to define the user profile is applied to the participants. The information collected with this questionnaire is: gender, age, area where the participant is living, class schedule, if currently is working, three activities that the participant enjoys outside the home, activities that the participant enjoys at home but without technology, time dedicated to study activities, if the participant has a cell phone for exclusive use, if the participant has pets; and the current level or semester at the university. Additionally, the RSI history questionnaire is submitted by each participant; this contributes to the generation of an integral user profile needed to accomplish the main goal of this project.

The demographic questionnaire captures average cellphone usage, which correlates with the continuous device use characteristics of software engineering tasks. Recent studies have shown that IT professionals, including software engineers, frequently use smartphones and other mobile devices for both work and personal purposes during working hours. This high level of device use is associated with increased work-life conflict and potential health risks, including musculoskeletal disorders and technostress [40], [41], [42].

Once the user profile information was collected, a pilot test was conducted using a main dataset created from 8 users, which was then used as input for the prompting model stage in this methodological process.

### C. Study design

This study adopts a within-subjects repeated measures design with two within-subject factors:

- LLM (five levels, corresponding to five different large language models).
- Time of day (two levels: morning and afternoon).

All participants will receive ten non-digital activity recommendations: five in the morning and five in the afternoon. Each set will include one recommendation per LLM. This design enables direct comparison of individual responses to different models while accounting for potential variations due to the time of day.

To evaluate each recommendation independently, participants are recorded on video while interacting with each other. Each video is temporally annotated to mark the start and end of each recommendation, enabling fine-grained alignment with the participant's facial expressions and emotional responses.

To minimize emotional carryover between different LLMs, a brief neutral pause (15–20 seconds) is introduced after each pair of recommendations, before the next model is presented. During this pause, the screen displays a black background along with a simple guided breathing instruction (e.g., "Take a slow, deep breath and relax"). This design aims to help participants reset their emotional state before engaging with recommendations from a different LLM.

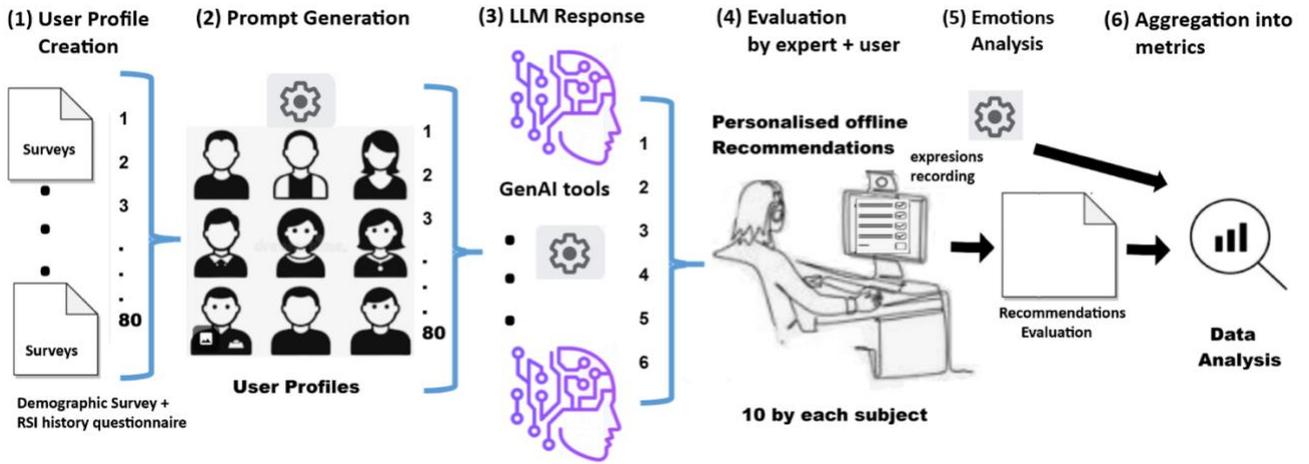

*Figure 1. Procedure of the experiment proposed.*

### D. Large Language Model Selection

The selection of the five different LLMs was based on the type of license of use (mainly open source) with high performance in related fields [43] and focused to perform well in personalized recommendation systems [44]. Therefore, the LLMs selected are:

Gemini (by Google DeepMind), Phi-4 (by Microsoft), Mistral (by Mistral AI), Qwen 2.5 (by Alibaba DAMO), LLaMA 3.2 (by Meta).

The LLMs used in this research are highly representative of the period under consideration, as no prior studies employing generative AI tools to provide non-digital recommendations in the RSI domain were identified in the related work survey. While newer versions of each LLM may be released depending on the pace of core development, they generally retain the same underlying algorithmic foundation across distributions and updates. Given that software engineering is a dynamic field, it is important to track studies within specific temporal contexts to generate new research insights and maintain relevance over time [45].

### E. Prompting model

As shown in Figure 1, when the user profiles are well-defined, the next step includes scenario-based prompt generation to be then applied in each GenAI tool. Prompts had to be slightly adjusted for participant profiles during a pilot test.

Since the recommendations provided by each GenAI model are intended to suggest non-technological activities within scenarios constrained by time and space, it is necessary to construct a user profile based on variables that capture both individual characteristics (e.g., preferences, health history) and environmental context (e.g., location, time availability). To this end, the previously mentioned demographic questionnaire meets these criteria, which are aligned with those used for prompt generation [46].

The construction of the prompt was guided by a modular methodology aligned with prior research on prompt design in mental health contexts [47]. The prompt was iteratively refined across five stages: (i) an initial Basic Prompt was implemented using a single instruction, which led to overly general, temporally inappropriate, and impractical recommendations; (ii) through systematic error detection, issues such as irrelevant activities, long outputs, or use of technology were identified, leading to the incorporation of temporal constraints and contextual relevance conditions; (iii) a User Profile module was added to personalize suggestions based on interests, preferences, and daily routines; (iv) dynamic variables such as time of day, weather, academic schedule, and geolocation were integrated to simulate real-time adaptation; and (v) a final constraint was applied to limit the response length to 40 words, optimizing for clarity and semantic precision. This multi-stage process follows the six-module decomposition strategy (Persona, Task, N-shot, Input, Output, Template) proposed in [47], ensuring a context-aware, user-aligned, and replicable prompting structure for fair model comparison.

Accordingly, the process of prompt generation is automated. A script is designed to extract each user profile from the main dataset. Based on time and space constraints, two distinct scenarios are generated: the first scenario begins with a randomized date and time within the interval of 00:00h to 11:59h. Given the corresponding geolocation and timestamp, it is possible to retrieve historical weather data, which becomes part of the context provided to each LLM. The second scenario follows a similar procedure, but the time interval ranges from 12:00 to 23:59h. All relevant variables are then syntactically integrated into the prompt model. The final structure of the prompt is as follows:

*"Generate a single, immediate activity that does not involve expenses, for the following profile: '+profile+' The activity must be aligned with their interests and help encourage a healthy disconnection from continuous cell phone use, through activities that promote well-being, health, and academic performance. Attention must be paid to the following conditions: '+weatherReport+' From Monday to Friday the student is in classes, you must pay attention to their shift (morning, afternoon, or night), so depending on the current time, the recommendation should be to stop using the phone and pay attention to their classes. The activity must be possible in Cuenca, Ecuador, and consistent with the current time. If it is a nighttime schedule between 11:00 PM and 4:00 AM, suggest that they sleep; if it is outside of that range, suggest activities at home. The student should have lunch if the current time is between 12:30 PM and 2:30 PM. Do not copy the current weather report verbatim or use nicknames in the generated text. It is mandatory to keep the total length of the generated text to around 40 words. Eliminate initial expressions such as \*\*actividad\*\* or similar. Also, use*

*neutral language appropriate to the Ecuadorian population (do not use the terms "parcero" or "parcera").*"

This adjusted prompt is then parsed into each selected tool, incorporating the user's profile data and environmental features. Each model generates two recommendations, resulting in ten new entries being added to the main dataset. As shown in Figure 2, this information is then ready for evaluation by health professionals and users, according to their respective responsibilities.

Since this is a framework proposal, and based on the theoretical background and related work, the main idea is to assess LLM-driven offline recommendations to evaluate their relevance in terms of user profile and context awareness. The results will provide insights from the user's questionnaires and indicate whether it will be necessary to add or remove specific features to test further algorithms in future work.

## V. RESULTS AND DATA ANALYSIS

A normality test will be applied to the data to analyze the data from the 10 daily recommendations for 80 final-year computer science students, integrating technical evaluation metrics (MCC-score, F1-score, Precision and Recall) and User Satisfaction. Depending on the result, a parametric or non-parametric test will be selected to analyze the data.

This section describes the planned analysis approach. No results are available yet, as data collection is part of the second stage of the registered report process.

## VI. THREATS TO VALIDITY

Several potential threats to validity have been identified in the planning phase of this study, and corresponding mitigation strategies will be implemented.

To preserve internal validity, the risk of order effects and participant fatigue is acknowledged, given that each participant will complete two sessions (morning and afternoon) within the same day. To mitigate this, the order of the LLM-generated recommendations will be counterbalanced or randomized, and short breaks between sessions will be allowed if needed. Additionally, different but comparable recommendations will be provided in each session to reduce potential learning or adaptation effects. To further control emotional carryover between models, a short-guided breathing pause after each LLM block and a brief neutralization phase at the start of the session are included. Additionally, to reduce potential bias in LLM-generated recommendations, prompt engineering strategies were employed to carefully configure queries directed to different generative AI systems, with the goal of generating more precise and contextually relevant results. Furthermore, the generated responses were subject to manual evaluation and refinement by subject matter experts to ensure that the final recommendations were accurate, trustworthy, and consistent with professional standards.

Regarding construct validity, the possibility of social desirability bias in self-reported satisfaction ratings is recognized. To address this, participant responses will remain anonymous, and objective emotional indicators (e.g., facial expression data) will be collected to complement the subjective assessments.

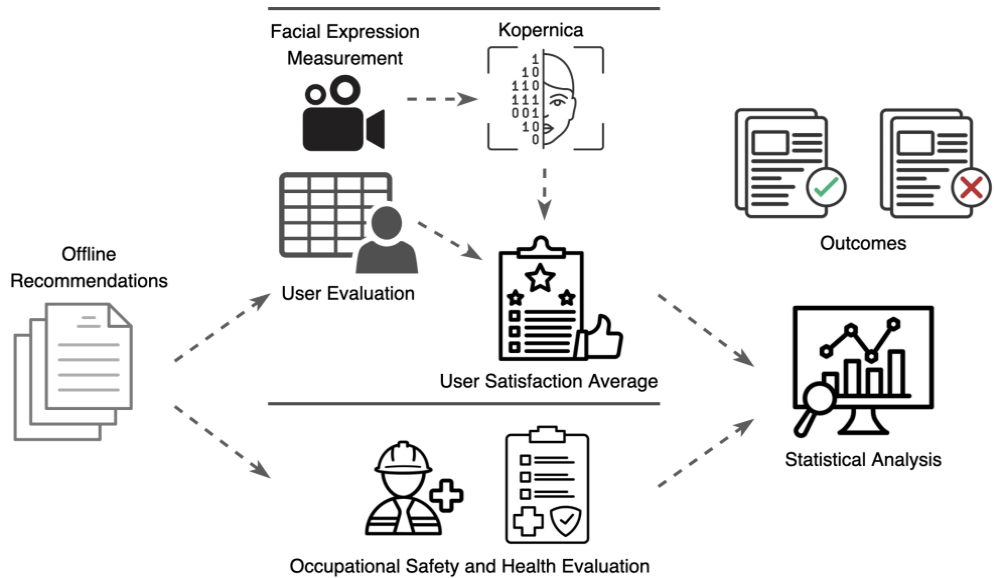

*Figure 2. Stages and interactions in the recommendations analysis.*

Concerning the self-report instrument for physical discomfort or RSI history brief, an adapted version of the NMQ questionnaire will be used to ensure consistent interpretation. As for the facial expression analysis, potential inaccuracies or cultural biases in emotion detection software are acknowledged; A commercial tool, Kopernica, developed by the Neurologyca company, will be used and the results will be interpreted alongside subjective feedback (e.g., SUS responses).

Regarding external validity, the study will rely on a relatively homogeneous sample 80 final-year students in Computer Science. While this limits the generalizability of findings to broader populations (e.g., older adults or non-technical users), it enhances ecological validity, as RSI is a well-documented concern among individuals in computing-related fields. As such, participants are likely to have relevant personal experience and be able to provide meaningful feedback on the perceived usefulness and relevance of the recommendations.

## VII. ETHICAL ISSUES

Ethical issues will be considered when the GenAI tools provide the recommendations and when the participants evaluate the personalized offline recommendations. Additionally, authorization o the Committee for the Evaluation of Interventions in Human Beings of the University of Cuenca will be obtained.

The participants will be clearly informed about the purpose of the study, the type of data collected (demographic information, satisfaction ratings, and facial expressions), and how their data will be used. Participation will be entirely voluntary and written informed consent by university bioethics committee will be taken before any data collection takes place (See Appendices Section).

To protect privacy and confidentiality, all data will be anonymized, and participants will not be identified in any report or publication. Facial expression analysis will be used solely for research purposes and interpreted alongside subjective feedback (e.g., SUS scores) to assess emotional responses. Only the emotion outputs (e.g., detected emotional states during the interaction with each recommendation) will be stored. No video or audio recordings of participants will be saved.

Given that the study evaluates the relevance of generative AI–based recommendations in the context of preventing RSI, it directly concerns participant health and safety. Therefore, all generated content will be manually reviewed by researchers and verified by an occupational health professional. Only recommendations deemed appropriate, safe, and non-harmful for the user profile will be presented during the study.

The data collected during the evaluation will be stored on a secure server at the University of Cuenca. They will be available for future use (e.g., replication of the evaluation) based on an agreement between the interested parties.

## VIII. REFERENCES

[1] V. C. Storey, W. T. Yue, J. L. Zhao, and R. Lukyanenko, "Generative Artificial Intelligence: Evolving Technology, Growing Societal Impact, and Opportunities for Information Systems Research," *Information Systems Frontiers*, pp. 1–22, Feb. 2025, doi: 10.1007/S10796-025-10581-7/TABLES/4.

[2] A. Said, "On explaining recommendations with Large Language Models: a review," *Front Big Data*, vol. 7, p. 1505284, Jan. 2024, doi: 10.3389/FDATA.2024.1505284/BIBTEX.

[3] M. Abdullah, "Effects of Repetitive Strain/Stress Injury on the Human Body," *International Journal of Medical and Health Sciences*, vol. 13, no. 12, pp. 494–500, Nov. 2019, doi: 10.5281/ZENODO.3593218.

[4] B. C. Marcoux, V. Krause, and E. R. Nieuwenhuijsen, "Effectiveness of an educational intervention to increase knowledge and reduce use of risky behaviors associated with cumulative trauma in office workers," *Work*, vol. 14, no. 2, pp. 127–135, 2000, Accessed: Jun. 21, 2025. [Online]. Available: https://pubmed.ncbi.nlm.nih.gov/12441528/

[5] J. P. Patel, K. P. Thekdi, and A. R. Gohel, "Study of musculoskeletal problems among long-term computer users in Ahmedabad City," *Natl J Physiol Pharm Pharmacol*, vol. 13, no. 5, pp. 1050–1050, May 2023, doi: 10.5455/NJPPP.2023.13.03133202326032023.

[6] "eTools : Computer Workstations - Work Process and Recognition | Occupational Safety and Health Administration." Accessed: Jun. 21, 2025. [Online]. Available: https://www.osha.gov/etools/computer-workstations/work-process

[7] A. de Waal, A. Killian, A. Gagela, J. Baartzes, and S. de Klerk, "Therapeutic Approaches for the Prevention of Upper Limb Repetitive Strain Injuries in Work-Related Computer Use: A Scoping Review," *J Occup Rehabil*, vol. 35, no. 2, pp. 234–267, Jun. 2024, doi: 10.1007/S10926-024-10204-Z/FIGURES/3.

[8] M. Monsey, I. Ioffe, A. Beatini, B. Lukey, A. Santiago, and A. B. James, "Increasing compliance with stretch breaks in computer users through reminder software," *WORK: A Journal of Prevention, Assessment & Rehabilitation*, vol. 21, no. 2, pp. 107–111, Jan. 2003, doi: 10.3233/WOR-2003-00312.

[9] L. Trujillo and X. Zeng, "Data entry workers perceptions and satisfaction response to the 'Stop and Stretch' software program," *WORK: A Journal of Prevention, Assessment & Rehabilitation*, vol. 27, no. 2, pp. 111–121, Jan. 2006, doi: 10.3233/WOR-2006-00554.

[10] R. A. Henning, P. Jacques, G. V. Kissel, A. B. Sullivan, and S. M. Alteras-Webb, "Frequent short rest breaks from computer work: effects on productivity and well-being at two field sites," *Ergonomics*, vol. 40, no. 1, pp. 78–91, Jan. 1997, doi: 10.1080/001401397188396.

[11] G. Nwatuzie and G. A. Nwatuzie, "HUMAN-CENTERED SOFTWARE ENGINEERING: ENHANCING DEVELOPER PRODUCTIVITY AND WELL-BEING," 2015. [Online]. Available: www.wjert.org

[12] P. Godliauskas, · Darja Šmite, F. Calefato, H. Khalajzadeh, and I. Steinmacher, "The well-being of software engineers: a systematic literature review and a theory," *Empirical Software Engineering 2024 30:1*, vol. 30, no. 1, pp. 1–42, Dec. 2024, doi: 10.1007/S10664-024-10543-8.

[13] M. Perera, S. Ananthanarayan, C. Goncu, and K. Marriott, "The Sky is the Limit: Understanding How Generative AI can Enhance Screen Reader Users' Experience with Productivity Applications," *Conference on Human Factors in Computing Systems - Proceedings*, Apr. 2025, doi: 10.1145/3706598.3713634/SUPPL_FILE/PN2039.PDF.

[14] F. Suni-Lopez and N. Condori-Fernandez, "Evaluation of an Emotion-Aware Persuasive Framework Based on Peripheral Interaction for Reducing Physical Strain in Office

[15] M. Chen et al., "Evaluating Large Language Models Trained on Code," Jul. 2021, Accessed: Jun. 21, 2025. [Online]. Available: http://arxiv.org/abs/2107.03374

[16] M. Imran and N. Almusharraf, "Google Gemini as a next generation AI educational tool: a review of emerging educational technology," *Smart Learning Environments*, vol. 11, no. 1, pp. 1–8, Dec. 2024, doi: 10.1186/S40561-024-00310-Z/METRICS.

[17] M. Abdin et al., "Phi-4 Technical Report," Dec. 2024, Accessed: Jun. 21, 2025. [Online]. Available: https://arxiv.org/abs/2412.08905v1

[18] M. Sasaki, N. Watanabe, and T. Komanaka, "Enhancing Contextual Understanding of Mistral LLM with External Knowledge Bases," Apr. 2024, doi: 10.21203/RS.3.RS-4215447/V1.

[19] O. Aydin, E. Karaarslan, F. S. Erenay, and N. Bacanin, "Generative AI in Academic Writing: A Comparison of DeepSeek, Qwen, ChatGPT, Gemini, Llama, Mistral, and Gemma," Feb. 2025, Accessed: Jun. 21, 2025. [Online]. Available: https://arxiv.org/abs/2503.04765v2

[20] M. Masalkhi et al., "A side-by-side evaluation of Llama 2 by meta with ChatGPT and its application in ophthalmology," *Eye 2024 38:10*, vol. 38, no. 10, pp. 1789–1792, Feb. 2024, doi: 10.1038/s41433-024-02972-y.

[21] D. Morris, J. B. Brush, and B. R. Meyers, "SuperBreak: Using interactivity to enhance ergonomic typing breaks," *Conference on Human Factors in Computing Systems - Proceedings*, pp. 1817–1826, 2008, doi: 10.1145/1357054.1357337.

[22] L. McLean, M. Tingley, R. N. Scott, and J. Rickards, "Computer terminal work and the benefit of microbreaks," *Appl Ergon*, vol. 32, no. 3, pp. 225–237, 2001, doi: 10.1016/S0003-6870(00)00071-5.

[23] "Effectiveness of Excersize Mini-Breaks - Articles - TIFAQ." Accessed: Jun. 21, 2025. [Online]. Available: https://www.tifaq.org/articles/excersize_mini-breaks-may98-arthur_saltzman.html

[24] M. O. Ayemowa, R. Ibrahim, and M. M. Khan, "Analysis of Recommender System Using Generative Artificial Intelligence: A Systematic Literature Review," *IEEE Access*, vol. 12, pp. 87742–87766, 2024, doi: 10.1109/ACCESS.2024.3416962.

[25] S. Sen, V. Gonzalez, E. J. Husom, S. Tverdal, S. Tokas, and S. O. Tjøsvoll, "ERG-AI: enhancing occupational ergonomics with uncertainty-aware ML and LLM feedback," *Applied Intelligence*, vol. 54, no. 23, pp. 12128–12155, Dec. 2024, doi: 10.1007/S10489-024-05796-1/FIGURES/10.

[26] C. Wohlin, P. Runeson, M. Höst, M. C. Ohlsson, B. Regnell, and A. Wesslén, "Experimentation in Software Engineering," *Springer Nature*, no. 2nd ed, 2024, doi: 10.1007/978-3-662-69306-3.

[27] R. van Solingen, V. Basili, G. Caldiera, and H. D. Rombach, "Goal Question Metric (GQM) Approach," *Encyclopedia of Software Engineering*, Jan. 2002, doi: 10.1002/0471028959.SOF142.

[28] G. Rau and Y. S. Shih, "Evaluation of Cohen's kappa and other measures of inter-rater agreement for genre analysis and other nominal data," *J Engl Acad Purp*, vol. 53, p. 101026, Sep. 2021, doi: 10.1016/J.JEAP.2021.101026.

[29] M.-S. Hosseini, F. Jahanshahlou, M. A. Akbarzadeh, M. Zarei, and Y. Vaez-Gharamaleki, "Formulating research questions for evidence-based studies," *Journal of Medicine, Surgery, and Public Health*, vol. 2, p. 100046, Apr. 2024, doi: 10.1016/J.GLMEDI.2023.100046.

[30] M. Hossin and Sulaiman, "A REVIEW ON EVALUATION METRICS FOR DATA CLASSIFICATION EVALUATIONS," *International Journal of Data Mining & Knowledge Management Process (IJDKP)*, vol. 5, no. 2, 2015, doi: 10.5121/ijdkp.2015.5201.

[31] D. Chicco and G. Jurman, "The advantages of the Matthews correlation coefficient (MCC) over F1 score and accuracy in binary classification evaluation," *BMC Genomics*, vol. 21, no. 1, pp. 1–13, Jan. 2020, doi: 10.1186/S12864-019-6413-7/TABLES/5.

[32] K. Wolf, "Measuring facial expression of emotion," *Dialogues Clin Neurosci*, vol. 17, no. 4, pp. 457–462, Dec. 2015, doi: 10.31887/DCNS.2015.17.4/KWOLF.

[33] P. Desmet, "Faces of product pleasure; 25 positive emotions in human-product interactions," *International Journal of Design*, vol. 6, no. 2, pp. 1–29, 2012, Accessed: Jun. 21, 2025. [Online]. Available: https://research.tudelft.nl/en/publications/faces-of-product-pleasure-25-positive-emotions-in-human-product-i

[34] P. Vlachogianni and N. Tselios, "Perceived usability evaluation of educational technology using the System Usability Scale (SUS): A systematic review," *Journal of Research on Technology in Education*, vol. 54, no. 3, pp. 392–409, 2022, doi: 10.1080/15391523.2020.1867938.

[35] "Ergonomics of human-system interaction-Human-centred design for interactive systems COPYRIGHT PROTECTED DOCUMENT," 2019, Accessed: Jun. 21, 2025. [Online]. Available: www.iso.org

[36] L. Mateos-González, J. Rodríguez-Suárez, J. A. Llosa, and E. Agulló-Tomás, "Versión española del Nordic Musculoskeletal Questionnaire: adaptación transcultural y validación en personal auxiliar de enfermería," *An Sist Sanit Navar*, vol. 47, no. 1, p. e1066, Mar. 2024, doi: 10.23938/ASSN.1066.

[37] M. Hossin and Sulaiman, "A Review on evaluation metrics for data classification evaluations," *International Journal of Data Mining & Knowledge Management Process (IJDKP)*, vol. 5, no. 2, 2015, doi: 10.5121/ijdkp.2015.5201.

[38] D. Chicco and G. Jurman, "The advantages of the Matthews correlation coefficient (MCC) over F1 score and accuracy in binary classification

evaluation," *BMC Genomics*, vol. 21, no. 1, Jan. 2020, doi: 10.1186/s12864-019-6413-7.

[39] T. A. E. Prasetya, N. I. A. Samad, A. Rahmania, D. A. Arifah, R. A. A. Rahma, and A. Al Mamun, "Workstation Risk Factors for Work-related Musculoskeletal Disorders Among IT Professionals in Indonesia," *Journal of Preventive Medicine and Public Health*, vol. 57, no. 5, pp. 451–460, Sep. 2024, doi: 10.3961/JPMPH.24.214.

[40] E. Whelan and O. Turel, "Personal use of smartphones in the workplace and work–life conflict: a natural quasi-experiment," *Internet Research*, vol. 34, no. 7, pp. 24–54, Dec. 2024, doi: 10.1108/INTR-08-2022-0607.

[41] H. Blake, J. Hassard, J. Singh, and K. Teoh, "Work-related smartphone use during off-job hours and work-life conflict: A scoping review," *PLOS Digital Health*, vol. 3, no. 7, p. e0000554, Jul. 2024, doi: 10.1371/journal.pdig.0000554.

[42] D. Lalloo, J. Lewsey, S. V Katikireddi, E. B. Macdonald, and E. Demou, "Health, lifestyle and occupational risks in Information Technology workers," *Occup Med (Chic Ill)*, vol. 71, no. 2, pp. 68–74, Apr. 2021, doi: 10.1093/occmed/kqaa222.

[43] S. Pagano *et al.*, "Evaluating ChatGPT, Gemini and other Large Language Models (LLMs) in orthopaedic diagnostics: A prospective clinical study," *Comput Struct Biotechnol J*, vol. 28, pp. 9–15, Jan. 2025, doi: 10.1016/j.csbj.2024.12.013.

[44] S. Chavan, S. Giri, S. Gornar, A. Kalsapnavar, R. Mirajkar, and A. Suryawanshi, "EduMate: AI-Based Student Support Platform with OCR and Voice Assistance," *2025 International Conference on Emerging Smart Computing and Informatics, ESCI 2025*, 2025, doi: 10.1109/ESCI63694.2025.10988345.

[45] M. Nejjar, L. Zacharias, F. Stiehle, and I. Weber, "LLMs for science: Usage for code generation and data analysis," *Journal of Software: Evolution and Process*, vol. 37, no. 1, p. e2723, Jan. 2025, doi: 10.1002/SMR.2723.

[46] S. Bang and H. Song, "LLM-based User Profile Management for Recommender System," Feb. 2025, Accessed: Jun. 12, 2025. [Online]. Available: http://arxiv.org/abs/2502.14541

[47] M. Kim, T. Kim, T. H. A. Vo, Y. Jung, and U. Lee, "Exploring Modular Prompt Design for Emotion and Mental Health Recognition," *Conference on Human Factors in Computing Systems - Proceedings*, p. 18, Apr. 2025, doi: 10.1145/3706598.3713888.

IX. APPENDICES

Additional information can be found in the following URL: https://tinyurl.com/ESEM25RR